\documentclass[smallextended]{svjour3}       
\usepackage{graphicx}
\usepackage{xcolor}

\begin{document}

\title{Long-term non-linear predictability of ENSO events over the 20th century}

\author{H. F. Astudillo \and R. Abarca-del-Rio \and  F. A. Borotto}

\institute{H. F. Astudillo and F. A. Borotto \at Departamento de F{\'i}sica, Universidad de Concepci{\'o}n, P.O. Box 160--C, Concepci{\'o}n, Chile
\and R. Abarca--del--Rio  \at Departamento de Geof{\'i}sica, Universidad de Concepci{\'o}n, P.O. Box 160--C, Concepci{\'o}n, Chile}

\maketitle  
 
\begin{abstract}
We show that the monthly recorded history (1878-2013) of the Southern Oscillation Index (SOI), a
 descriptor of the El Ni\~no Southern Oscillation (ENSO) phenomenon, can be well described as a 
dynamic system that supports an average nonlinear predictability well beyond the spring barrier. 
The predictability is strongly linked to a detailed knowledge of the topology of the attractor obtained by embedding the SOI index in a wavelets base state space. Using the state orbits on the
attractor we show that the information contained in the Southern Oscillation Index (SOI) is sufficient to provide average nonlinear predictions for time periods of 2,  3 and 4 years in advance
throughout the 20th century with an acceptable error. The simplicity of implementation and 
ease of use makes it suitable for studying non linear predictability in 
any area where observations are similar to those that describe the ENSO phenomenon.

\end{abstract}

\section{Introduction}
The El Ni\~no Southern Oscillation is a large scale tropical Pacific atmosphere-ocean phenomena.
\cite{Zebiak1987},  \cite{Neelin1998}, \cite{Wang2012} one of the stronger quasi-oscillatory pattern observed in
the climate system that induces important changes in the global Earth's gravity field \cite{Phillips2012},
influence precise variations of geodetics parameters such as Geocenter \cite{Cretaux2002} may play a major role in
the
generation of seismicity \cite{Gillas2010}, or serve as a modulator of the external sun's influence on Earth's
climate \cite{Burn2013}. However teleconnections not only influences climate worldwide but also
influences economic, political and social related aspects of the Earth system  \cite{Bjerknes1969}, \cite{Kovats2003}, 
\cite{Brunner2002}, \cite{Hsiang2011}. Thus, ENSO signal is a key index representing the complex dynamics of the Earth
system as a whole \cite{McPhaden2006}. It is therefore understandable that  comprehension of its dynamics is one of the
most important scientific milestones today and in parallel its prediction is one of the major challenges, and no longer
an element solely of interest to climatologists alone.

International efforts to forecast ENSO using both dynamical and statistical methods, have met with some success. Thus,
with lead times up to 6 mo, various forecasts performed reasonably well, whereas for longer lead times, that is
passing through the boreal spring barrier, the performance becomes rather low, particularly for the most extreme events
such as El Ni\~no 1982-83 and 1997-98's \cite{Landsea2000}, \cite{Chen2008}. However recent results
\cite{Ludescher11022014} highlight potential skill in predicting tropical Pacific variability at lead times
exceeding 1 year for some events \cite{Fedorov}. This is an evidence that the ability to predict ENSO events for
periods longer than one year and perhaps particularly those leading to greater consequences worldwide is also feasible.
However, the previous stage before prediction, is to first show  that this goal is achievable. That is, there is at
least enough information within the oscillatory system, as represented by the main ENSO index, the Southern Oscillation
Index (SOI), that may allow prediction of ENSO events.

The main purpose of this paper is to report that such evidence exists. In a precedent article \cite{hfa2010}, a
methodology for applying Takens's embedding theorem \cite{Takens1981} to reconstruct an event is developed, by taking
into account just the local information contained in a short time series. The physical basis of the method lies in the
definition of the local predictability \cite{Ding1}. Local predictability limit gives a measure of long time-scale local
predictability on the attractor \cite{Ding2}. These results outlined the capabilities of the methodology which this
local reconstruction method uses to characterize the
local and non-linear properties of a given system dynamics. Thus,
with the knowledge of the variability over the past 20th century, it was possible to reconstruct local
extreme ENSO events (1982-83 and 1997-98), leading by more than two years in advance. Likewise, the method can be
applied to arbitrarily large sizes but the search for the relevant embedding state, or - space subspaces that
characterize the phenomena requires a high computational cost, hindering the use of the method, and thereby reproduction
of results. A simpler but complementary approach allows ease reproducibility, allowing to reconfirm the precedent study.
Thanks to a significant breakthrough that overcame the computational and technical difficulties involved in the
application of the method to embed a signal of limited temporal extension in a state space. As \cite{wavelets2011} it
has already been reported that the wavelet space and the embedding space are equivalent in topology, therefore a
description is possible  by using under Takens's embedding theorem, which allows to reconstruct  a state space which
is diffeomorphic to the physical phase space from a time series \cite{RevModPhys.65.1331}.  
Thus, in this analysis, the SOI is considered as a macroscopic variable that has been embedded in the $n$-dimensional
state space. So that, we assume that the SOI represents a variable that is a part of non-linear differential equations
describing the oscillation. Extending  \cite{hfa2010} to the problem under consideration, a wavelet decomposition
separates the different $n$ state coordinates from the signal equivalently  as described in \cite{PhysRevLett.45.712}
and \cite{PhysRevLett.59.845}. Thus liberating us from the tedious searching process for the optimal state spaces as was
done previously. Thus, the process's presented in the precedent investigation can be eased, making it more attractive
and straightforward for its application and reproducibility.

Thus, the methodology presented here allows to search whithin the whole length of the times series non linear trajectories that do present the same properties that the non
 linear trajectories of the attractor of a particular event investigated. Therein it is possible using
 the sole non-linear information already present whitin a time series to find similar attractor-trajectories that allows 
to investigate if a particular ENSO event is reproducible to a certain extent.  

\section{Methodology}
To achieve this goal we first define an n-dimensional state space $\mathbf{E_n} $. The evolution of the system in this state space is recorded in a state trajectory given by 
$ \mathbf{r}_n (t) = (S_1 (t), ..., S_n (t)) $, where quantities $ S_1 (t),..., S_n(t)$ are 
obtained by wavelet decomposition of the SOI signal \cite{Konnen1998}.  From the original signal, 
a given number $n$ of  component signals $S_i$ are obtained by means of wavelet filtering 
the monthly SOI index between the years 1866-2013. We decompose the SOI data using a Discrete Wavelet
 Transformation (DWT) multiresolution algorithm  with a Morlet mother wavelet function 
\cite{Craigmile2002}, \cite{Murguia2011}.  In our case we choose an 8 level
decomposition, considering the monthly time series resolution. Because fewer levels did not allow to separate efficiently the interannual time scales or capture interdecadal time scales efficiently  
that characterize ENSO \cite{Wang2012}, so each of the components ($S_i$, $i$ from 1 to 8) represent a given bandwidth (1-2 intra-seasonal, 3-5 interannual, 6-8, decadal to interdecadal) whereas the last
 component $S_9$ represents lowest frequencies, that describes the trend. In other words 
$SOI= S_1+...+S_9$. In this paper we use a state space $\mathbf{E_9} $ following \cite{Tsonis2009} encrypting all information contained in the SOI in the state trajectory, but we also define
a six dimensional subspace,
$\mathbf{E_6}$ where the state trajectory is given by $\mathbf{r_6} (t) = (S_4 (t), ..., S_9 (t))$ containing information only in the low frequency range (lower than $1/yr)$. We compute the low 
frequency part of the SOI signal as $S_L=S_4+...+S_9$.

\begin{figure*}
\vspace*{2mm}
\par
\begin{center}
\includegraphics[width=14cm]{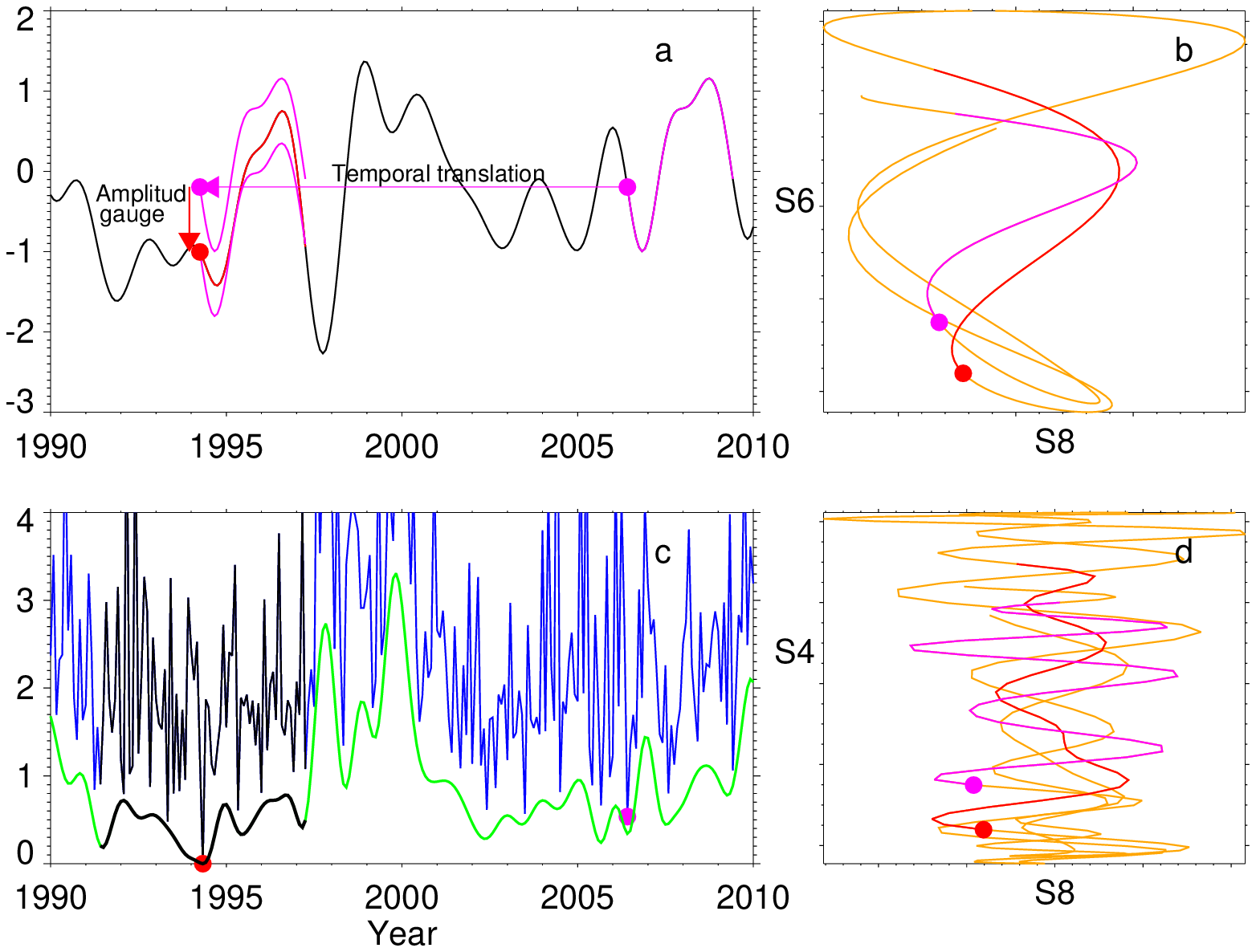}
\end{center}
\caption{Figure details the steps to be carried out to determine which part of the signal $S_L$ that is most similar to
the event to be predicted. In panel \textbf{a} the $S_L$ signal is plotted. The red dot is the starting point of
the event that corresponds to the first starting state point in Figure \ref{figure1}, while the red section indicates
the
extent of the event to be predicted. The magenta point indicates the point corresponding to the nearest point on the
chosen orbit. In panel \textbf{c} the blue line is the Euclidean distance between the starting point and each point of
the state trajectory calculated in $\mathbf{E_9}$, while the green curve is the Euclidean distance between the starting
point and each point of the path that was calculated in $\mathbf{E_6} $. In panels \textbf{b} and \textbf{d} we plot
$(S_8,S_6)$ and $(S_8,S_4)$ two-dimensional projections of the attractor, respectively. The orange curves are
projections of the state trajectory.}
\label{figure0}
\end{figure*}

The event (An El Ni\~no or La Ni\~na event) to be predicted  $S_{L_i}$ is selected from $S_L$ (represented by the red
segment in Figure \ref{figure0}a).  The first point in $S_{L_i}$ (represented by a red dot in Figure \ref{figure0}a) is
the starting point at time $t_{sp}$ and the terminal point is located by $T$ steps in the future.  To select another
part of the signal $S_L$ that enables to reproduce the relevant features of the $S_{L_i}$ event, we seek a state point
of the state trajectory in the attractor around which the embedded signal $SOI$ has the same topological features like
than around the starting point of the event. Thus, both state points are located about the same region in the attractor.
Figure \ref{figure0}b and Figure \ref{figure0}d shows bidimensional projections of the state trajectory (in orange)
around the attractor, while red and magenta colors plots shows the selected event and the most similar event,
respectively. To accomplish this task we calculate the Euclidean distance of all state points in the state trajectory to
the starting state point. In the $\mathbf{E_9} $ reconstructed state space the Euclidean distance between each state
point of state trajectory to the starting state point is given by $d(t)= \parallel \mathbf{r_9}(t) -
\mathbf{r_9}(t_{sp}) \parallel$, which is plotted in blue in Figure \ref{figure0}c. Then, we collect the times, $t_{j}$,
where the curve $d = d(t)$ exhibits local minima. 
Each selected local minimum of the function $d(t)$ corresponds to a state point on each orbit that is closest to the
starting state point in the reconstructed state space $\mathbf{E_9}$. This search is performed away from a region
of $T$ steps from each side (to the past and the future) of the starting point, $t_{sp}$ (regions in black in Figure
\ref{figure0}c). For each of the collected times, $t_{j}$ a piece of the corresponding $S_L$ of $T$ steps long  is
extracted into its own future (that is $S_{L_j}$). As shown in Figure \ref{figure0}a, the $S_{L_j}$ piece (magenta) is
then moved to the starting point, $t_{sp}$ (the red dot) through a temporal translation (the magenta arrow). In
addition, the translated signal (the $S_{L_j}$ piece),  as an amplitud gauge (the red arrow in Figure \ref{figure0}), is
shifted up or down so that its first point (originally at $t_{j}$) matches the value of $S_L$ at the starting point at
time $t_{sp}$ (the event to be predicted, that is $S_{L_i}$). In addition, the physical consistency implies that all the
selected $S_{L_j}$ pieces must contain the same number of maxima and minima than the chosen event in $S_L$ ($S_{L_i}$).
Finally, of all the $S_{L_j}$ pieces selected and translated as described, the piece that contains the same number of
maxima and the same number of minima as $S_{L_i}$  which produces the least quadratic difference is definitely chosen
(the magenta dot in Figure \ref{figure0}).  

\begin{figure}
\vspace*{2mm}
\par
\begin{center}
\includegraphics[width=8.3cm]{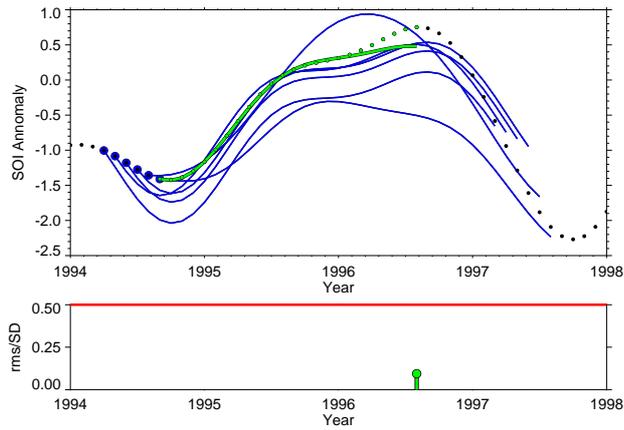}
\end{center}
\caption{Above: The Figure outlines the steps followed in the procedure for obtaining an average prediction for two
years. Points in black are the monthly amplitudes of low-frequency SOI index ($S_L$). The blue points are the starting
points for each prediction for three years, which are drawn with blue lines.The green curve is the two years average
nonlinear prediction. Below: The red line marks the boundary where the mean square error of the
average nonlinear prediction is half the standard deviation of $S_L$. The green bar at the bottom indicates the mean
square error relative to the standard deviation of $S_L$ of the average nonlinear prediction drawn in green.}
\label{figure1}
\end{figure}

To make a robust estimate of the event we perform an average of six consecutive months of predictions resulting a
$T_{eff}= T-12$ months prediction, which is plotted in green in Figure \ref{figure1}, after a necessary second amplitude
gauge. Thus, the averaged prediction begins on the date of the last prediction used for constructing the 6 monthly mean
that has the final date of the first prediction used. Thus on average an initial three years local prediction reduces to
an mean prediction of about two years.

\begin{figure}
\vspace*{2mm}
\par
\begin{center}
\includegraphics[width=8.3cm]{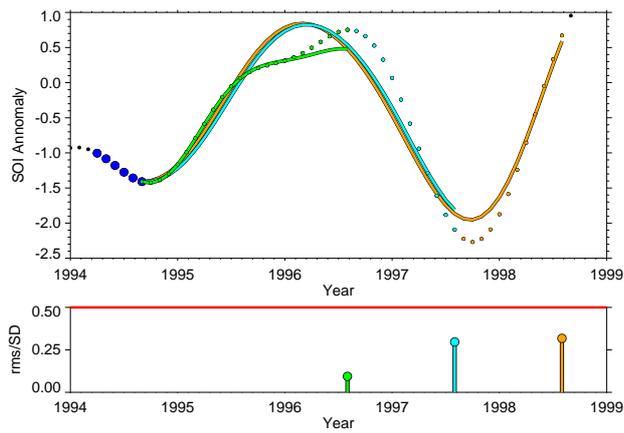}
\end{center}
\caption{ Above: Average non-linear predictions for two years lead (in green) and the $S_L$ time series (in black) are
drawn.  Below: The relative error. The red line indicates the $0.5$ threshold. The green, cyan, and orange bars are
the average non linear predictions error for $T_{eff} =2, 3$ and
$4$ years, respectively.}
\label{figure2}
\end{figure}

Finally we compute the error of the prediction of these two year long prediction by normalizing to the standard deviation of $S_L$. The error is reported as a bar below the signal. The red line in
Figures \ref{figure1} to \ref{figure4} indicate the threshold of $0.5$ which means that the error of the prediction reach the half of the standard deviation
($SD$) of the whole signal $S_L$.

\section{Results}

\begin{figure*}
\vspace*{2mm}
\par
\begin{center}
\includegraphics[width=14cm]{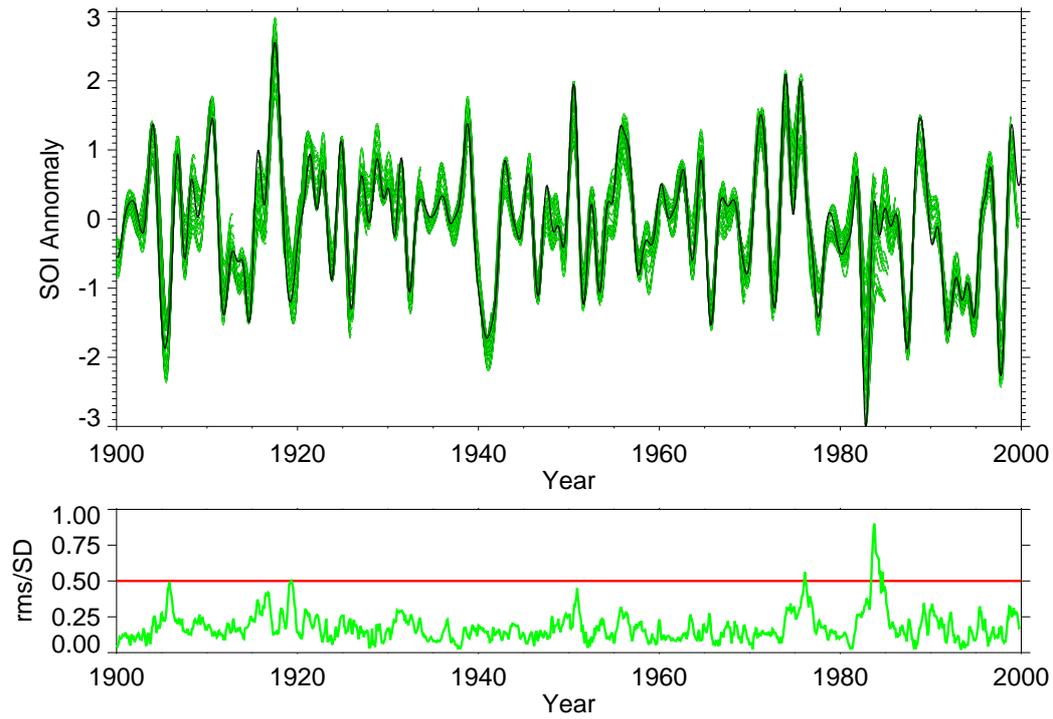}
\end{center}
\caption{ Above: The performance of the procedure for average non-linear predictions with $T_{eff} = 2$ years of El
Ni{\~n}o during the 20th century. Below: The relative error.  The red line indicates the $0.5$ threshold.}
\label{figure3}
\end{figure*}

Remarkably (Figures \ref{figure1}, \ref{figure2} and \ref{figure3}) one of the most extreme event of the 20th century, namely the 1997-98 Ni\~no event could be clearly anticipated well beyond a two
year lead to the least. Thus it is possible to find trajectories in the SOI time series that present the same nonlinear characteristics than the variability of the 1997-98 event. Thus, the relative
error for the two year lead Figure
\ref{figure3}b is below the 0.5 threshold. Instead for the 1982-83 event, the relative error is well above the 0.5 threshold. That is, meaning that it is not possible to find any correct
trajectory/attractor within the whole 1878-2013 time span that present the same non-linear properties than the 1982-83 event.

The analysis of relative error histograms (Figure \ref{figure4}a) shows that when the lead times increases (from 2 to 4
years), the averaged relative error indeed increases, from $0.177$, $0.298$ and $0.389$ for $T_{eff} =2, 3, 4$ years
respectively (see Figure \ref{figure4}a and Table \ref{tabla1}).

Indeed while increasing the lead times, the error for the particular 1982-83 and 1997-98 events increases. However, for
the El  Ni{\~n}o 1997-98, while increasing the lead times, the relative error always falls below the 0.5 threshold (lesser than 0.37).
The orbit/trajectories that are closer to an event and that allows the accurate averaged prediction, changes depending
on lead time (and can change from one month to another). However, the analysis of Figure \ref{figure4}b shows that the
histogram of closer trajectories is a Gaussian-shaped distribution, that is, in general the non-linear prediction
always finds a trajectory in the vicinity of the event in the state space.

Thus, if we had to look for the trajectories used for the construction of the 2 year lead 
prediction of the particular 1997-98 event, throughout the span of the event, these would be those of the 1911, 1972-73
and 1982-83 events. Interestingly the orbits/trajectories that naturally emerges for the 1982-83 event, and eventhough this event rises above the 0.5
threshold, are indeed preferentially the 1997-98 (in the future) and also the 1901-02 events.

\section{Discussion and Conclusions}

As  already stated the physical support of this methodology consider the dynamics of ENSO
described as a dynamic system. Thus, the signal, can be considered as a solution for a nonlinear system
of differential equations describing the oscillation. In virtue of the Takens's theorem the signal is embedded in the
$9$-dimensional state space $\mathbf{E_9} $ so that the relevant information contained in the system is encrypted in a
state trajectory of the system. With this procedure the state points of the state trajectory, ie, the states of the
system are arranged in orbits around an attractor.
The idea behind this methodology relies on possibility to find in past or future evolutions of a time series events that present the same characteristics (orbits around the attractor) than the event investigated. The closest orbits increases the capacity of event predictability. 

Once an event $S_{L_i}$ is selected in the signal  $S_L$, the starting point at time $t_{sp}$ and a time interval of
prediction $T$ is determined. The starting point of the selected event corresponds to a system state point on the
attractor. To find an event in the signal, $S_{L_j}$, corresponding to a similar solution of the same set of deferential
equations, we determine the set of state points $t_{sp_j}$ that are in a region around the starting state point in the
attractor (in $\mathbf{E_9} $). We remark that the reconstructed state space $\mathbf{E_9} $ is diffeomorphic to the
physical phase space \cite{RevModPhys.65.1331}.

The selection of the most similar piece of the event signal is by performing the choicest among all segments of length
$T$, starting at state points that are in vicinity of the starting state point of the event. These datas provide the
least deviation standard with the same number of peaks and the same number of minima as the event signal.

\begin{figure}
\vspace*{2mm}
\par
\begin{center}
\includegraphics[width=9cm]{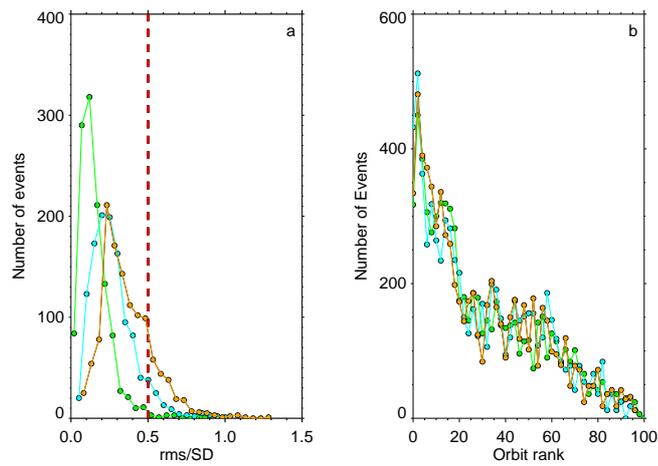}
\end{center}
\caption{ a) Histograms of the relative errors of the average non-linear predictions of El Ni{\~n}o during
the 20th century are shown. Relative errors for 2, 3 and 4 years are depicted in green, cyan and orange colors,
respectively. b) the histograms of the rank of the selected orbit
for each non-linear prediction is been represented. The histogram of orbit rank for 2, 3 and 4 years are represented
using green, cyan and orange colors, respectively.}
\label{figure4}
\end{figure}

The most important result of this study is that the so-called SOI anomaly corresponds to the dynamics of a nolinear
oscillator having complex regularities and exhibits an acceptable level of accuracy of average non-linear predictability
in the range between 2 and 4 years of time span.
Although the topology of the attractor is unknown, for each event it is always possible to find an orbit corresponding
to another event that ocurs in the past or in the future of the starting point which has  the same category.
The knowledge of the topological structure of the attractor is a necessary task for forecasting. In our method, as
discussed previously, we chose the best orbit in the attractor. The histograms at the bottom of Figures \ref{figure4}a
and \ref{figure4}b shows that there is no deterministic rule that allows us to choose the right orbit with certainty,
although a clear trend towards the nearest orbit was observed. 

Note that indeed the methodology here used, as it uses a wavelet filter bank with different
 scales does take in account, when constructing the orbit attractor for a particular event, a part 
of the closest (past or future) evolution of the time series. However, this is performed only for constructing the attractor and search 
for similitudes within the whole length of the time series and not for constructing the prediction. 
 Therefore, this lapse is already existing and is not built in  any way with the method. In addition,
 the methodology find trajectories that are far away in time from the event analysed.
Also, in a precedent paper \cite{hfa2010} we already demonstrated that the methodology
 worked although search for relevant embedding space was tedious and with high computer cost. Here 
we only extend the methodology, thanks to the wavelet space to a set of embedding state spaces that indeed facilitates the search for closest orbits and therefore for similar
 trajectories within the length of the time series.  In other words the
 wavelet filtering bank does not interfere with the time selected. Finally the method is not filter
 dependent. Thus, we tested the methodology with two different filtering banks, that is with Empirical
 modal decomposition EMD \cite{Huang2005} and Singular Spectrum analysis SSA \cite{Golyandina2013}. In both cases, and particularly for SSA, 
the methodology could reconstruct correctly most events throughout the 20th century but failed for the 1982/83 event.

Although we do not know exactly the selection rule to determine the optimal orbits that allows forecasting that does not
invalidate the fact that the SOI signal has built enough information to reproduce events over at least a 2 years span, surpassing throughout the 20th century, with an acceptable error, with exception
of the 1982-1983 event.

\begin{acknowledgements}
This work was partially supported by Departamento de F\'isica and Departamento de Geof\'isica, Universidad de 
Concepci\'on, Concepci\'on, Chile.
We thank Dr. M. Paulraj for useful comments, criticism, and advice. 
The Southern Oscillation Index (SOI, from CRU (Climate Research Unit, UK)) was obtained from the climate explorer web site
(http://climexp.knmi.nl).
\end{acknowledgements}

\begin{table}[b]
\caption{Mean, Variance, Skweness and Kurtosis of normalized errors distributions
shown in Figures \ref{figure4}a and \ref{figure4}b. }\label{tabla1}

\begin{tabular}{@{\vrule height 10.5pt depth4pt  width0pt}c c c c c c }
\hline
\textbf{$T_{eff}$} & \textbf{$E_n$} & \textbf{M}& \textbf{V} & \textbf{S} & \textbf{K}\\
\hline
2 & $E_9$ & 0.177 & 0.011 & 2.145 & 8.056 \\
\hline
3 & $E_9$ & 0.298 & 0.020 & 1.404 & 3.090 \\
\hline
4 & $E_9$ & 0.389 & 0.028 & 1.147 & 2.067 \\
\hline
2 & $E_6$ & 0.235 & 0.019 & 1.825 & 4.927 \\
\hline
3 & $E_6$ & 0.379 & 0.035 & 1.236 & 2.338\\
\hline
4 & $E_6$ & 0.483 & 0.069 & 4.080 & 29.49\\
\hline
\end{tabular}
\end{table}

\end{document}